\documentclass[aps,prl,showpacs,showkeys,amsmath,amssymb,
twocolumn,
floatfix,
superscriptaddress
]{revtex4-1}
\usepackage{graphicx}
\usepackage{times}
\usepackage{verbatim}
\usepackage{color}
\usepackage{bm}
\usepackage{natbib}
\usepackage[normalem]{ulem}
\definecolor{gray}{rgb}{0.05,0.05,0.55}
\usepackage[colorlinks,linkcolor=gray,anchorcolor=gray,citecolor=gray, urlcolor=gray,filecolor=gray]{hyperref}

\begin{document}

\title{Comment on ``Comment on Daya Bay's definition and use of $\Delta m^2_{ee}$'' by S. Parke and R. Zukanovich Funchal}

\newcommand{\ECUST}{\affiliation{Institute of Modern Physics, East China University of Science and Technology, Shanghai}}
\newcommand{\IHEP}{\affiliation{Institute~of~High~Energy~Physics, Beijing}}
\newcommand{\Wisconsin}{\affiliation{University~of~Wisconsin, Madison, Wisconsin 53706}}
\newcommand{\Yale}{\affiliation{Wright~Laboratory and Department~of~Physics, Yale~University, New~Haven, Connecticut 06520}} 
\newcommand{\BNL}{\affiliation{Brookhaven~National~Laboratory, Upton, New York 11973}}
\newcommand{\NTU}{\affiliation{Department of Physics, National~Taiwan~University, Taipei}}
\newcommand{\NUU}{\affiliation{National~United~University, Miao-Li}}
\newcommand{\Dubna}{\affiliation{Joint~Institute~for~Nuclear~Research, Dubna, Moscow~Region}}
\newcommand{\CalTech}{\affiliation{California~Institute~of~Technology, Pasadena, California 91125}}
\newcommand{\CUHK}{\affiliation{Chinese~University~of~Hong~Kong, Hong~Kong}}
\newcommand{\NCTU}{\affiliation{Institute~of~Physics, National~Chiao-Tung~University, Hsinchu}}
\newcommand{\NJU}{\affiliation{Nanjing~University, Nanjing}}
\newcommand{\TsingHua}{\affiliation{Department~of~Engineering~Physics, Tsinghua~University, Beijing}}
\newcommand{\SZU}{\affiliation{Shenzhen~University, Shenzhen}}
\newcommand{\NCEPU}{\affiliation{North~China~Electric~Power~University, Beijing}}
\newcommand{\Siena}{\affiliation{Siena~College, Loudonville, New York  12211}}
\newcommand{\IIT}{\affiliation{Department of Physics, Illinois~Institute~of~Technology, Chicago, Illinois  60616}}
\newcommand{\LBNL}{\affiliation{Lawrence~Berkeley~National~Laboratory, Berkeley, California 94720}}
\newcommand{\UIUC}{\affiliation{Department of Physics, University~of~Illinois~at~Urbana-Champaign, Urbana, Illinois 61801}}
\newcommand{\SJTU}{\affiliation{Department of Physics and Astronomy, Shanghai Jiao Tong University, Shanghai Laboratory for Particle Physics and Cosmology, Shanghai}}
\newcommand{\BNU}{\affiliation{Beijing~Normal~University, Beijing}}
\newcommand{\WM}{\affiliation{College~of~William~and~Mary, Williamsburg, Virginia  23187}}
\newcommand{\Princeton}{\affiliation{Joseph Henry Laboratories, Princeton~University, Princeton, New~Jersey 08544}}
\newcommand{\VirginiaTech}{\affiliation{Center for Neutrino Physics, Virginia~Tech, Blacksburg, Virginia  24061}}
\newcommand{\CIAE}{\affiliation{China~Institute~of~Atomic~Energy, Beijing}}
\newcommand{\SDU}{\affiliation{Shandong~University, Jinan}}
\newcommand{\NanKai}{\affiliation{School of Physics, Nankai~University, Tianjin}}
\newcommand{\UC}{\affiliation{Department of Physics, University~of~Cincinnati, Cincinnati, Ohio 45221}}
\newcommand{\DGUT}{\affiliation{Dongguan~University~of~Technology, Dongguan}}
\newcommand{\XJTU}{\affiliation{Department of Nuclear Science and Technology, School of Energy and Power Engineering, Xi'an Jiaotong University, Xi'an}}
\newcommand{\UCB}{\affiliation{Department of Physics, University~of~California, Berkeley, California  94720}}
\newcommand{\HKU}{\affiliation{Department of Physics, The~University~of~Hong~Kong, Pokfulam, Hong~Kong}}
\newcommand{\UH}{\affiliation{Department of Physics, University~of~Houston, Houston, Texas  77204}}
\newcommand{\Charles}{\affiliation{Charles~University, Faculty~of~Mathematics~and~Physics, Prague}} 
\newcommand{\USTC}{\affiliation{University~of~Science~and~Technology~of~China, Hefei}}
\newcommand{\TempleUniversity}{\affiliation{Department~of~Physics, College~of~Science~and~Technology, Temple~University, Philadelphia, Pennsylvania  19122}}
\newcommand{\CUC}{\affiliation{Instituto de F\'isica, Pontificia Universidad Cat\'olica de Chile, Santiago}} 
\newcommand{\CGNPG}{\affiliation{China General Nuclear Power Group, Shenzhen}}
\newcommand{\NUDT}{\affiliation{College of Electronic Science and Engineering, National University of Defense Technology, Changsha}} 
\newcommand{\IowaState}{\affiliation{Iowa~State~University, Ames, Iowa  50011}}
\newcommand{\ZSU}{\affiliation{Sun Yat-Sen (Zhongshan) University, Guangzhou}}
\newcommand{\CQU}{\affiliation{Chongqing University, Chongqing}} 
\newcommand{\BCC}{\altaffiliation[Now at ]{Department of Chemistry and Chemical Technology, Bronx Community College, Bronx, New York  10453}} 

\newcommand{\UCI}{\affiliation{Department of Physics and Astronomy, University of California, Irvine, California 92697}} 
\author{D.~Adey}\IHEP
\author{F.~P.~An}\ECUST
\author{A.~B.~Balantekin}\Wisconsin
\author{H.~R.~Band}\Yale
\author{M.~Bishai}\BNL
\author{S.~Blyth}\NTU
\author{D.~Cao}\NJU
\author{G.~F.~Cao}\IHEP
\author{J.~Cao}\IHEP
\author{J.~F.~Chang}\IHEP
% ...

\collaboration{The Daya Bay Collaboration}\noaffiliation
\date{\today}
\maketitle

``Comment on Daya Bay's definition and use of $\Delta m^2_{ee}$'' by S. Parke and R. Zukanovich Funchal~\cite{PZ} seems to have confounded two different concepts: an experimental measurement vs. the interpretation of the measurement. We clarify a few points in our response.

First, all relevant Daya Bay publications~\cite{DB2014,DB2015,DB2017,DB2018} have consistently reported two values of $\Delta m^2_{32}$ in the standard three-neutrino framework under the assumption of the normal or inverted mass hierarchy. These values were always obtained through a fit with the exact full three-neutrino oscillation formula and used the best knowledge of the solar oscillation parameters at the time.

Second, in all these publications we also reported the value of $\Delta m^2_{ee}$ through another fit, independent from the one mentioned above, with the formula
\begin{eqnarray}
P_{ee}=1 &-& \sin^2\left(2\theta_{13}\right) \sin^2\left(\Delta m_{ee}^{2} \frac{L}{4E}\right)  \nonumber\\
          &-& \cos^4\theta_{13} \sin^2\left(2 \theta_{12}\right) \sin^2\left(\Delta m_{21}^{2} \frac{L}{4E}\right) \,.
\end{eqnarray}
Such a fit is viable since a reactor neutrino experiment at kilometer baselines is only sensitive to two effective neutrino oscillation frequencies: one leading frequency (instead of two) with an amplitude driven by $\theta_{13}$, and one sub-leading frequency with an amplitude driven by $\theta_{12}$. The leading frequency $\Delta m^2_{ee}$, naturally a constant as a fitting parameter, is our measurement. It enables interpretation in various theoretical models, either in the three-neutrino framework or beyond.

The main motivation for the use of $\Delta m^2_{ee}$ is to report our observations in a model-independent way. In Eq.~1 at Daya Bay's baseline, the sub-leading oscillation has been well measured by KamLAND~\cite{KamLAND} and the leading oscillation is well supported by our data. Therefore, we fit $\Delta m^2_{ee}$ based on existing experimental facts, largely independent of the three-neutrino framework. Our measurement $\Delta m^2_{ee}$ does not depend on the choice of mass ordering. Moreover, it can be interpreted in other models and/or as new measurements come to light.

In the supplemental material of Ref.~\cite{DB2015}, we provided a discussion about the interpretation of this quantity in the three-neutrino framework. In this supplement, two interpretations were provided: one with a slight L/E dependence, which was identified as the second Daya Bay definition $\Delta m^2_{ee}(\mathrm{DB2})$  in Parke and Zukanovich Funchal's comment, and another one with a constant $\Delta m_\phi^2=5.17\times 10^{-5}$ eV$^2$ offset between $\Delta m^2_{ee}$ and $\Delta m^2_{32}$. Both were demonstrated to be numerically equivalent to the one proposed by Nunokawa, Parke and Zukanovich Funchal in Ref.~\cite{npz} across Daya Bay's L/E regime in the three-neutrino model.

It is important to note that we have never defined $\Delta m^2_{ee}$ in terms of a combination of fundamental oscillation parameters. Instead, all measurements of $\Delta m^2_{ee}$ reported to date by Daya Bay used the effective oscillation model of Eq.~1 as the primary definition of this parameter without exception. When we fit the data, $\Delta m^2_{ee}$ is an independent parameter. It is not necessary nor advantageous to impose an additional relation with the fundamental parameters. To avoid confusion, recent Daya Bay publications have used the ``$\approx$'' sign instead of the ``='' in Eq.~1; however, this change has no impact on our fitting process nor on the interpretation of the parameter.

As a final comment, introducing the definition $\Delta m_{e e}^{2}(\mathrm{NPZ}) \equiv \cos ^{2} \theta_{12} \Delta m_{31}^{2}+\sin ^{2} \theta_{12} \Delta m_{32}^{2}$ as argued by Parke and Zukanovich Funchal~\cite{PZ}, 1) does not provide any new information since we have provided the fit value of $\Delta m^2_{32}$; 2) does not extend the approximate oscillation formula to the large L/E regime as all other similar interpretations, although itself is L/E independent; 3) would invalidate our publications in case new physics beyond the three-neutrino framework is found, e.g. the sterile neutrino.

Because of all these reasons, we disagree with the authors' criticism ``... Daya Bay's new definition of $\Delta m^2_{ee}$ does not manifestly show the simple relationship to the fundamental parameters of the neutrino sector for short baseline reactor experiments, such as Daya Bay and RENO. Nor is it useful for future medium baseline experiments like JUNO ...''. Daya Bay's reported value of $\Delta m^2_{ee}$ is a model-independent measurement of the leading oscillation frequency observed in our experiment. This definition is simple, intuitive, and supported by experimental observations.

In conclusion, Daya Bay will continue to extract $\Delta m^2_{ee}$ as a model-independent fitting parameter and to provide it to the community alongside the fundamental parameter $\Delta m^2_{32}$, obtained independently using the exact formula in the three-neutrino oscillation framework.  For experiments where the two-constant-frequency approximation does not apply, such as JUNO~\cite{JUNO}, the exact three-neutrino oscillation formula that explicitly depends on $\Delta m^2_{31}$  and $\Delta m^2_{32}$  should always be used.


\begin{thebibliography}{00}



\bibitem{PZ} S.~J.~Parke and R. Zukanovich Funchal, arXiv:1903.00148.

\bibitem{DB2014} F.~P.~An {\it et al.} (Daya Bay Collaboration), Phys.\ Rev.\ Lett.\ {\bf 112}, 061801 (2014).

\bibitem{DB2015} F.~P.~An {\it et al.} (Daya Bay Collaboration), Phys.\ Rev.\ Lett.\ {\bf 115}, 111802 (2015).

\bibitem{DB2017} F.~P.~An {\it et al.} (Daya Bay Collaboration), Phys.\ Rev.\ D\ {\bf 95}, 072006 (2017).

\bibitem{DB2018} D.~Adey {\it et al.} (Daya Bay Collaboration), Phys.\ Rev.\ Lett.\ {\bf 121}, 241805 (2018).

\bibitem{KamLAND} S.~Abe {\it et al.} (KamLAND Collaboration), Phys.\ Rev.\ Lett.\ {\bf 100}, 221803 (2008).

\bibitem{npz} H. Nunokawa, S. J. Parke and R. Zukanovich, Phys.\ Rev.\ D\ {\bf 72}, 013009 (2005).

\bibitem{JUNO}  F.~P.~An {\it et al.}, J. Phys. G: Nucl. Part. Phys. {\bf 43}, 030401 (2016).

\end{thebibliography}
\end{document}